\documentclass{article}%
\usepackage{amsmath}%
\usepackage{amsfonts}%
\usepackage{amssymb}%
\usepackage{graphicx}
\providecommand{\U}[1]{\protect\rule{.1in}{.1in}}

\begin{document}

\begin{center}
{\large On the application of fracture mechanics mixed-mode models of sliding
with friction and adhesion}

\bigskip

M.Ciavarella

Politecnico di BARI. DMMM dept. V Orabona, 4, 70126 Bari.

Hamburg University of Technology, Dep Mech Eng, Am Schwarzenberg-Campus 1,
21073 Hamburg, Germany

email: mciava@poliba.it
\end{center}

\bigskip Abstract: As recently suggested in an interestring and stimulating
paper by Menga, Carbone \&\ Dini (MCD), applying fracture mechanics energy
concepts for the case of a sliding adhesive contact, imposing also the shear
stress is constant at the interface and equal to a material constant (as it
seems in experiments), leads to a increase of contact area which instead is
never observed. We add that the rigorous MCD theory also predict a size effect
and hence a distortion of the JKR curve during sliding which is also not
observed in experiments. Finally, a simpler example with the pure mode I
contact case, leads in the\ MCD theory to an unbounded contact area, which is
difficult to interpret, rather than a perhaps more correct limit of the
Maugis-Dugdale solution for the adhesive sphere when Tabor parameter is zero,
that is DMT's solution. We discuss therefore the implications of the MCD
theory, although they may be rather academic: recent semi-empirical models,
with an appropriate choice of the empirical parameters, seem more promising
and robust in modelling actual experiments.

\bigskip{} \bigskip{} \textbf{Keywords: }Adhesion, JKR model, friction, soft
matter, fracture mechanics, mixed mode, cohesive models.

\bigskip{}

\section{Introduction}

In fracture mechanics, it is well known that mixed mode enhances the toughness
observed in pure mode I (Evans \& Hutchinson, 1989) due to crack faces
interlocking and friction resulting for roughness. Unfortunately, these
mixed-mode models are not physical laws or general energy principles, but are
intrinsically empirical. They are mostly of the form including a mode-mixity
function $f\left(  \psi\right)  $ (Hutchinson \& Suo, 1992) giving the
"effective toughness" $G_{c,eff}$ as%
\begin{equation}
G_{c,eff}=G_{Ic}f\left(  \psi\right)  \label{Gpsi}%
\end{equation}
where $G_{Ic}$ is mode I toughness (or surface energy, if we assume Griffith's
concept). Also, $\psi$ is phase angle
\begin{equation}
\psi=\arctan\left(  \frac{K_{II}}{K_{I}}\right)
\end{equation}
being $K_{II}$ and $K_{I}$ respectively the mode II and mode I stress
intensity factors. Cao and Evans (1989) experimentally looked at epoxy-glass
bimaterial interface, and in general various models for microscopic phenomena
affect the interface toughness, such as friction, plasticity and dislocation
emission (Hutchinson, 1990).

\bigskip When models like these are applied to contact mechanics problem in
the presence of adhesion and friction, we may expect either the contact area
to be largely unaffected by \textit{the presence of a mode II loading}, in one
limit or that $G_{c}\simeq G_{Ic}$, and in this case we effectively expect
mode II \textit{weakens} the mode I condition, so the contact area should
decrease in sliding with respect to the pure adhesion case. The case of area
enhancement is rather unexpected, as experimentally it is confirmed
(Ciavarella, 2018, Papangelo \& Ciavarella, 2019, Sahli \textit{et al.} 2019,
Papangelo\textit{ et al.} 2019).

Fracture mechanics concepts were firstly applied by Johnson, Kendall and
Roberts (JKR-theory, 1971) to adhesion between elastic bodies, are applicable
to contact even in the presence of friction, as mixed-mode fracture mechanics
problem, but with some special peculiarities. Specifically, the energetic
"JKR-assumptions" correspond to the Griffith criterion, which consists in
practice in assuming extremely short range adhesive forces (virtually a
delta-function), the correct limit for soft and large bodies, and the
equivalent to the so called "small-yield" criterion is expressed by the Tabor
parameter (Tabor, 1977) \ for the sphere,
\begin{equation}
\mu_{sphere}=\left(  \frac{RG_{Ic}^{2}}{E^{\ast2}\Delta r^{3}}\right)
^{1/3}=\frac{\left(  Rl_{a}^{2}\right)  ^{1/3}}{\Delta r}=\frac{\sigma_{0}%
}{E^{\ast}}\left(  \frac{R}{l_{a}}\right)  ^{1/3} \label{Tabor}%
\end{equation}
where $R$ is the sphere radius, $G_{Ic}$ is work of adhesion, $\Delta r$ is
the range of attraction of adhesive forces, and $E^{\ast}$ the plane strain
elastic modulus. $E^{\ast}=\left(  \frac{1-\nu_{1}^{2}}{E_{1}}+\frac{1-\nu
_{2}^{2}}{E_{2}}\right)  ^{-1}$ and $E_{i},$ $\nu_{i}$ are the Young modulus
and Poisson ratio of the material couple. Also, $\sigma_{0}$ is the
theoretical strength of the material, and we have introduced the length
$l_{a}=G_{Ic}/E^{\ast}$ as an alternative measure of adhesion --- for
Lennard-Jones potential of elastic crystals, $l_{a}\simeq0.05a_{0}$, where
$a_{0}$ is the equilibrium distance, which means that $l_{a}$ is of the order
of angstroms.

The use of energetic criteria extending JKR to the presence of friction was
attempted by Savkoor and Briggs (1977) who also conducted experiments between
glass and rubber. Writing the energy balance condition in terms of a constant
tangential load for which $K_{II}=Y\tau_{m}\sqrt{\pi a}$, where $\tau_{m}$ is
the average shear and $X,Y$ are geometric factors (which contain also some
averaging over the perimeter, also of the $K_{III}$ term), while $a$ is the
radius of the circular contact area, Savkoor and Briggs's results in a reduced
effective energy for the "ideally brittle fracture" at equilibrium
\begin{equation}
G_{c}=G_{Ic}=G_{I}+G_{II}=X^{2}\frac{p_{m}^{2}\pi a}{E^{\ast}}-Y^{2}\frac
{\tau_{m}^{2}\pi a}{E^{\ast}}=\left(  X^{2}\frac{p_{m}^{2}}{E^{\ast}}%
-Y^{2}\frac{\tau_{m}^{2}}{E^{\ast}}\right)  \pi a
\end{equation}
and $\ $therefore the contact area will follow the JKR equation, but following
the equation
\begin{equation}
G_{Ic,eff}=G_{Ic}-Y^{2}\frac{\tau_{m}^{2}\pi a}{E^{\ast}} \label{Savkoor}%
\end{equation}
where $\tau_{m}$ is the friction average stress (in the limit case, we could
even consider in sliding as a first approximation, as we are not modelling the
details of the shear stress distributions). Experiments clearly evidenced a
\textit{reduction of the contact area} when tangential load was applied, but
less than expected from assuming $G_{c}\simeq G_{Ic}$, so less than the
prediction (\ref{Savkoor}). Experimental findings showed some interference
with the development of Schallamach wave which tend to permit slip without
affecting the contact.

More recent experiments continue to confirm contact area reduction at both
macroscopic and even smaller scales (Sahli \textit{et al.} 2019). Johnson
(1996, 1997) and Waters and Guduru (2009) have proposed different models to
take into account the interplay between two fracture modes, namely I and II.
In particular, Johnson (1997) attempts also to model slip explicitly with
cohesive models (as well as the mode I corresponding part), and even in this
case, the conclusion remains of the contact area reduction. Various recent
other papers (Ciavarella, 2018, Papangelo \& Ciavarella, 2019, Sahli
\textit{et al.} 2019, Papangelo\textit{ et al.} 2019) have shown that the size
and even the elliptical shape of the contact under shear are reasonably found
by these LEFM\ models over a wide range of loads and geometrical features,
despite the mixed-mode function strictly requires a complex functional form to
replicate faithfully the results and hence empirical fitting at least over one
set of results.\ Also, they suggested there is no obvious advantage in trying
to model the slip displacements (which correspond to recur to a cohesive
model, in the context of fracture mechanics), since this effect is essentially
included in the mixed-mode function. Obviously, in the empirical functions
models, one could use different empirical functions, and therefore be able to
model also enhancements of contact area size.

\bigskip

\section{The area enhancement MCD theory}

\bigskip A recent paper by Menga \textit{et al.} (2018) (MCD, in the
following), stemming from some experimental evidence that shear stress should
be constant at the interface during sliding at least in rubber vs glass (see
Chateauminois \& Fretigny 2008, but also MCD reference list), introduce an
interesting and stimulating variant of the friction and adhesion problem
suggesting an \textit{increase of contact area in sliding} --- even without
the need to postulate an increase of the mixed mode fracture energy. This
result would seem in contrast with present experimental evidence on the
contact area, but has the advantage to emerge naturally from an apparently
thermodynamic rigorous theory, although the range of validity should be
discussed. To reconcile the predictions with the experiments, MCD argue that
fluctuation in the stresses reduce the effective surface energy, or $G_{Ic}$.
Notice immediately that assuming the shear stresses are constant at the
interface, is equivalent to a fully cohesive model, i.e. a cohesive model like
suggested by Dugdale-Barenblatt and Maugis (see Maugis, 2013), when the size
of the cohesive zone fully extends to the entire "crack" (i.e. contact area).
While energy criteria can be still applied with cohesive models, which have
been devised to extend LEFM, this is the true limit case which is to be
treated with great care.

\bigskip Indeed, we could define a Tabor parameter for the shear problem
\begin{equation}
\mu_{sphere}^{shear}=\frac{\tau_{0}}{E^{\ast}}\left(  \frac{R}{l_{a}^{shear}%
}\right)  ^{1/3}%
\end{equation}
where $\tau_{0}$ is now a theoretical strength under shear, and in principle,
we introduced also a different adhesive length scale $l_{a}^{shear}$. It is
hard to estimate $\mu_{sphere}^{shear}$ exactly, but if we assume that
$l_{a}^{shear}\simeq l_{a}$, it is possible that $\tau_{0}<<\sigma_{0}$ and
hence as%
\begin{equation}
\mu_{sphere}^{shear}<<\mu_{sphere}%
\end{equation}
there is a region for which we can be in the intermediate range for which we
can apply essentially LEFM criteria in mode I and a fully cohesive model for
mode II, as implicitly, the authors of MCD theory assume. However, this leads
to non-obvious results, which pose a number of interesting questions worth examining.

Consider a system in which there is an uncoupled mode I problem (pressures,
normal displacements), and a tangential one (shear stresses, shear
displacements), as for example in the case of a rigid body against an
elastomer with $\nu=0.5$ (or, more generally, Dundurs' second constant equal
to zero). We can write the strain energy stored in the body as
\begin{equation}
U=\frac{1}{2}\int_{A}\sigma vdA+\frac{1}{2}\tau_{0}W
\end{equation}
where $A$ the contact area, $\sigma$ is the normal pressure and $v$ the normal
component of displacement, $W$ is the displaced volume in tangential direction
$W=Aw$ where $w$ is the mean tangential displacement which we can write as
$w=k\tau_{0}A^{1/2}/E^{\ast}$, where $k$ is a constant factor of the order of
1, which is not important here. Hence, we can write $\tau_{0}=\frac{wE^{\ast}%
}{kA^{1/2}}$ , $W=k\tau_{0}A^{3/2}/E^{\ast}$, $W=wA$, and $\frac{1}{2}\tau
_{0}W=\frac{1}{2}\frac{E^{\ast}A^{1/2}}{k}w^{2}$. \ In the classical
formulation of this problem in mixed mode fracture, the state variables would
be ($v,w,A$), and one would need to apply a Griffith energy minimum principle,
the condition
\begin{equation}
\left(  \frac{\partial U}{\partial A}\right)  _{v,w}=G_{Ic}%
\end{equation}
with imposed displacements $v,w$. This would not satisfy the requirement that
shear stress distribution is constant in the contact area: the solution has
the well-known LEFM square root singularities, in both the normal pressure
distribution, and also in the shear stresses. Indeed, it is the full stick
solution which incidentally is used with success in previous papers
(Ciavarella, 2018, Papangelo \& Ciavarella, 2019, Sahli \textit{et al.} 2019,
Papangelo\textit{ et al.} 2019), even for this problem under transition to
sliding, but using "empirical" mixed-mode functions to take into account of
the complex effect of the influence of mode II into mode I.

\bigskip In MCD, the authors explore what a pure energy condition implies
without any mixed-mode empirical function. By applying the Legendre transform
to change the state variables from ($v,w,A$) to ($v,\tau_{0},A$), the
trivially obtain a new thermodynamic potential $H$,
\begin{align}
H  &  =U-\left(  \frac{\partial U}{\partial W}\right)  _{v,A}W=U-\left(
\frac{\partial U}{\partial w}\frac{\partial w}{\partial W}\right)
_{v,A}W=U-\tau_{0}W\\
&  =\frac{1}{2}\int_{A}\sigma udA-\frac{1}{2}\tau_{0}W
\end{align}
as imposing the condition
\begin{equation}
\left(  \frac{\partial H}{\partial A}\right)  _{v,\tau_{0}}=G_{Ic}%
\end{equation}
leads to the solution under mode I displacement control, and mode II
"strength" control.

This leads to their eqt.26 (both under force, or under displacement control in
mode I)
\begin{equation}
G_{c,eff}=G_{Ic}+\Delta G_{Ic}\left(  a\right)  =G_{Ic}+\frac{4\tau_{0}^{2}%
a}{\pi E^{\ast}}\quad\label{menga}%
\end{equation}
i.e. the effective surface energy (or toughness of the interface) is
increased, rather than decreased in Savkoor's theory (\ref{Savkoor}), and
curiously of a very similar quantity, perhaps even exactly the same since
$\tau_{0}=\tau_{m}$, if $Y=\frac{2}{\pi}$. This should not be confused from
the result of the theories using the mode-mixity functions of the type
(\ref{Gpsi}), like (Ciavarella, 2018, Papangelo \& Ciavarella, 2019, Sahli
\textit{et al.} 2019, Papangelo\textit{ et al.} 2019), since in the latter
case, there is no size-effect associated to the contact area $a$, as there is
in (\ref{menga}), and this has profoundly \textit{different} implications, as
we shall explore.

\bigskip Indeed, the first reactions to this result are that

\begin{itemize}
\item (i) even in the limit of no surface energy $G_{Ic}\rightarrow0$, there
would be an "effective adhesion", as $G_{c,eff}\rightarrow\frac{4\tau_{0}%
^{2}a}{\pi E^{\ast}}$, which, under sufficiently large compressive normal
forces, would imply an \textit{unbounded increase }of the effective energy;

\item (ii) this equation also implies a \textit{distortion} of the JKR
solution load vs area of contact which instead in the AFM experiments by
Carpick\textit{ et al.}[13], was a nearly perfect fit, even during sliding,
leading to the conclusion that the force was simply proportional to the
contact area.
\end{itemize}

Therefore, this area enhancement theory is more difficult to reconcile with
experiments than models like (Ciavarella, 2018, Papangelo \& Ciavarella, 2019,
Sahli \textit{et al.} 2019, Papangelo\textit{ et al.} 2019), which of course
remains semi-empirical, but also are more robust in the predictions -
particularly not having any size effect in the increase of toughness.

The authors of the MCD theory recur to suggesting how large pressure
oscillations may compensate for this effect, but these are not an outcome of
the theory, and this should be further investigated. However, there seem to be
simpler explanations as to why the implied effects are not measured, as we
shall explain in the next paragraph with a simpler example, which leads to a
even more surprising conclusion.

\begin{center}

$%
\begin{array}
[c]{cc}%
{\includegraphics[
height=3.4346in,
width=6.7596in
]%
{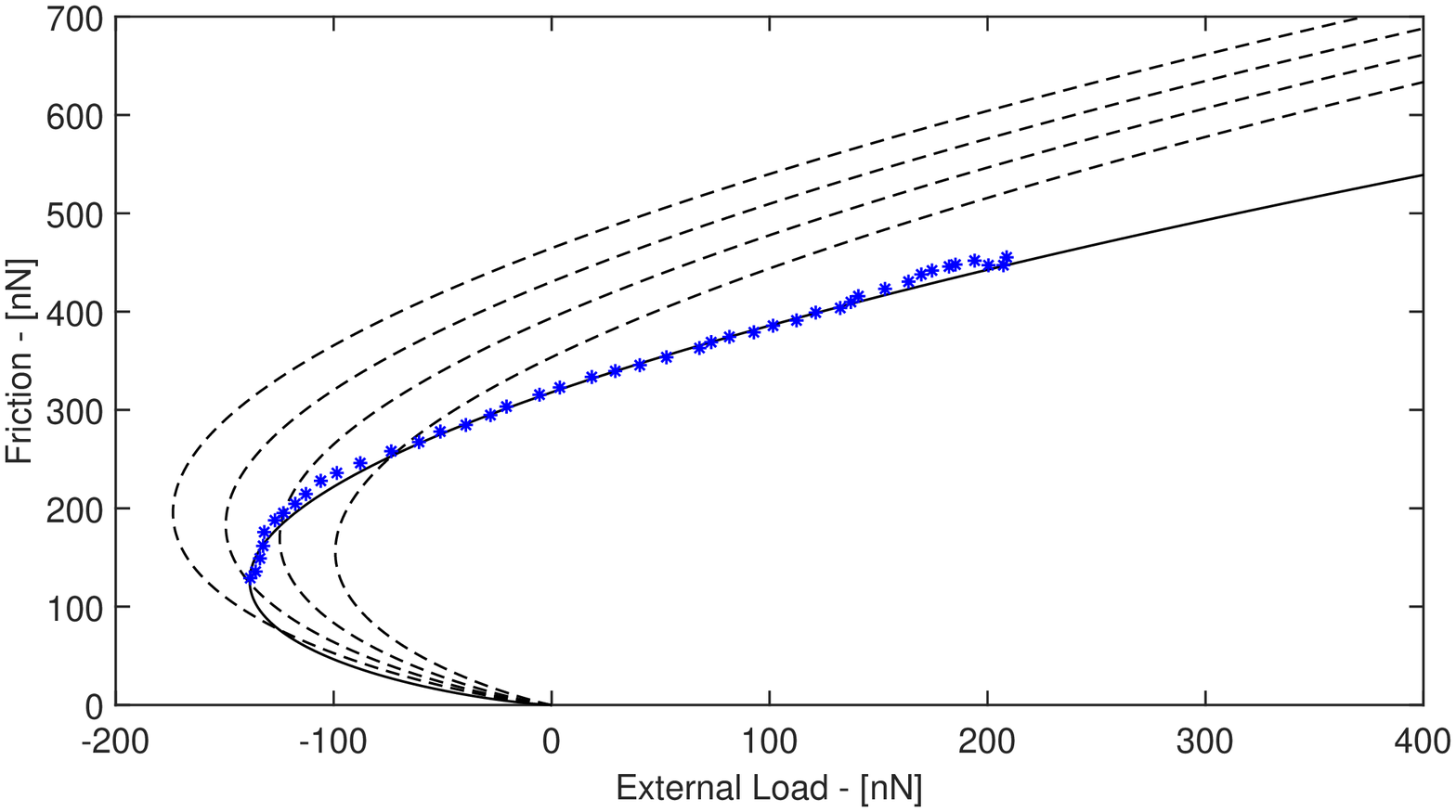}%
}
&
\end{array}
$

Fig.1. Friction measurements in UHV in the AFM (Carpick \textit{et al.} 1996).
Fitting the JKR area--load relationship (solid line) gives an extremely
accurate fit, which gives $G_{Ic}=0.21J/m^{2}$ and a uniform frictional stress
$\tau_{0}=0.84GPa$. The fit with eqt.26 of MCD theory is attempted with
$G_{Ic}=$ 0.01,0.04, 0.07 and 0.10 $J/m^{2}$ (dashed lines)
\end{center}

\section{\bigskip A mode I "nanoscale" paradox?}

\bigskip The MCD theory that the contact area should increase upon application
of tangential force may be due to their use of a energy condition for Linear
Elastic Fracrure mechanics, while being in the limit of fully developed
cohesive zones in mode II. \ If we now \bigskip speculate of an imaginary
adhesive problem in which "constant tensile pressure $\sigma_{0}$" is observed
the way MCD suggest, we have the strain energy is more simply
\begin{equation}
U=\frac{1}{2}\sigma_{0}V
\end{equation}
where $V$ is the displaced volume $V=Av$,\ where $v$ is the mean normal
displacement which we can write as $v=k\sigma_{0}A^{1/2}/E^{\ast}$, and as
usual $k$ is a constant factor of the order of 1, which is not important here.
Hence, $\sigma_{0}=\frac{vE^{\ast}}{kA^{1/2}}$ , $V=k\sigma_{0}A^{3/2}%
/E^{\ast}$, $V=vA$, and finally $U=\frac{1}{2}\frac{E^{\ast}}{k}A^{1/2}v^{2}$.

By applying the Legendre transform to change the state variables from ($v,A$)
to ($\sigma_{0},A$) in order to have the "pressure-control", we obtain a new
thermodynamic potential $H$,
\begin{align}
H  &  =U-\left(  \frac{\partial U}{\partial V}\right)  _{v,A}V=\frac{1}%
{2}\sigma_{0}V-\left(  \frac{\partial U}{\partial v}\frac{\partial v}{\partial
V}\right)  _{v,A}V\\
&  =-\frac{1}{2}\sigma_{0}V=-\frac{k}{2}\frac{\sigma_{0}^{2}}{E^{\ast}}A^{3/2}%
\end{align}
as indeed $-\sigma_{0}V$ is the potential energy associated with the uniform
stress distribution $\sigma_{0}$. Notice that now $v$ is no longer prescribed,
but the condition
\begin{equation}
\left(  \frac{\partial H}{\partial A}\right)  _{\sigma_{0}}=G_{Ic}
\label{conclusion}%
\end{equation}
is \textit{never} satisfied, as the energy decreases without limit and the
minimum is clearly \textit{at infinite contact area}. It is a similar
conclusion than MCD theory --- to the extreme limits the increase is unbounded.

In fact, this example instead may not be such an ideal situation at all: it is
known that at nanoscale, the adhesion problem becomes controlled by the
theoretical strength $\sigma_{0}$ (Gao \&\ Yao, 2004), and if we have a flat
indenter, the correct adhesion pull-off is simply
\begin{equation}
P_{po}=\sigma_{0}A
\end{equation}
where $A$ is the size of\ (flat ended) indenter, a principle which is found in
biology of nanoscale fibrillar structures to maximize their adhesion on their
feet, below sizes on the order of 100 nm. In the case of an indenter having a
shape, including the spherical classical one, one would need very small sizes
to reach this limit (Greenwood, 2009) unless the shape is the "optimal one"
involving elliptic integrals. In any case, the energetic formulation
simplified from MCD does not seem to provide a meaningful result, as
(\ref{conclusion}) seem to imply always infinite contact area, regardless of
shape of indenter. It is a limit case whose range of validity we are not able
to identify, but suggests a warning also on the more general mixed-mode
corresponding result.

\section{Discussion: the Dugdale-Maugis solution}

We have suggested that some hidden problems in the otherwise rigorous
thermodynamic theory of MCD may be due to the uncertain range of
applicability. Let us summarize what a well known cohesive model obtains in
the context of contact mechanics, for the mode I problem, namely the
Dugdale-Maugis solution for the contact problem of a sphere (Maugis, 2013).
The idea is to postulate a cohesive zone having an outer constant stress
$\sigma_{0}$ in an annulus $a<r<c$ outside of the contact, where $d=c-a$ is
the size of the cohesive zone. With this "trick", energetic methods can still
be applied to the problem, even beyond the LEFM formulation, since we know the
energy release rate of the cohesive zones. However, in the limit of very low
Tabor parameter, when the cohesive zone is extremely large, $m=c/a\rightarrow
\infty$, is very subtle, since the cohesive stresses are constant in the
annulus, but they must be zero, i.e. $\sigma_{0}=0$. A limit solution, the
so-called DMT-M solution for the sphere (Maugis's version of the DMT solution,
see Ciavarella (2017)), it doesn't appear possible to obtain it with the MCD
procedure, despite there is no reason why the same Legendre transform idea
could be applied in this simpler problem.

Hence, although we share with the authors of MCD theory the fascination for
the elegance of the Legendre pure thermodynamic formulation, and despite we
really enjoyed it for the number of stimulating discussion it generates, we
suggest it is problematic for various reasons:

\begin{itemize}
\item mixed mode problems have hardly been solved with simple energy
formulations without additional "empirical" constants and criteria, which is
why Hutchinson (1990) and the other references given in the introduction
paragraph, devised them for the problem of mixed mode fracture;

\item with respects to the empirical formulations which have found some
validation in experiments, it seems that MCD theory leads to a size-effect of
the surface energy/toughness which is not just giving an area increase, as MCD
noticed in the paper, but also contrasts with the excellent fits of JKR
theories done by Carpick \textit{et al} (1996) and many others;

\item there is a problem associated with the concept of an experiment
\textit{neither} under force \textit{nor} under displacement control, but
rather in "shear stress" control. As we have to decide how we are going to
cause the body to move, and the only ways we can postulate a proper problem
are (i) pushing at a constant force, (ii) pushing at a constant speed or (iii)
an intermediate case where a spring is connected to the body the end of which
is pushed at constant speed, it is hard to imagine an experiment (possibly
even a numerical one), that corresponds to the assumed conditions of the
thermodynamic Legendre transform theory in MCD's theory;

\item a fully cohesive developed zone in mode II, in the spirit of fracture
mechanics, corresponds to a very low Tabor parameter in shear, while the
energetic treatment can treat at most intermediate Tabor conditions. It is
hard to imagine a rigorous case in which the Tabor parameter in shear should
be extremely low, while the Tabor parameter in pressure should be very high.

\item similarly, trying to predict the contact area changes in a JKR\ (short
-range) theory due to a very long-range adhesion effect under shear appears
also possibly a problem.
\end{itemize}

\section{Conclusion}

The thermodynamics treatment of Menga \textit{et al.} (MCD theory) has
suggested a very interesting simple way to deal with the interaction of
adhesion and friction, which however leads to an increase of contact area
which, as noticed by the authors themselves, has not been observed in
experiments. We have additionally noticed that other paradoxical results
emerge which are contradicted in practice, namely a size effect of the surface
energy would distort the JKR curve nicely observed in many experiments, and an
application to a simpler purely adhesive problem leading to prediction of
infinite contact area. Although experimentalists have indeed been measured
directly in soft materials that shear stresses appear constant during sliding,
specifically in glass vs rubber, more recent deconvolutions considering the
high strain gradients reached (Nguyen, \textit{et al} 2011) start to find
deviations from the perfect constant shear distribution; this may be another
way out of the embarrassing results. We have discussed at length various
implications to the interesting MCD theory.

\section{\bigskip References}

Cao, H. C., \& Evans, A. G. (1989). An experimental study of the fracture
resistance of bimaterial interfaces. Mechanics of materials, 7(4), 295-304.

Ciavarella, M. (2017). On the use of DMT approximations in adhesive contacts,
with remarks on random rough contacts. Tribology International, 114, 445-449.

\bigskip Ciavarella, M. (2018). Fracture mechanics simple calculations to
explain small reduction of the real contact area under shear. Facta
universitatis, series: mechanical engineering, 16(1), 87-91.

Carpick, R. W., Agrait, N., Ogletree, D. F. \& Salmeron, M. 1996 Variation of
the interfacial shear strength and adhesion of a nanometer sized contact.
Langmuir 12, 3334--3340.

Chateauminois, A., \& Fretigny, C. (2008). Local friction at a sliding
interface between an elastomer and a rigid spherical probe. The European
Physical Journal E: Soft Matter and Biological Physics, 27(2), 221-227.

Evans, A. G., \& Hutchinson, J. W. (1989). Effects of non-planarity on the
mixed mode fracture resistance of bimaterial interfaces. Acta Metallurgica,
37(3), 909-916.

Gao, H., \& Yao, H. (2004). Shape insensitive optimal adhesion of nanoscale
fibrillar structures. Proceedings of the National Academy of Sciences,
101(21), 7851-7856.

Greenwood, J. A. (2009). Adhesion of small spheres. Philosophical Magazine,
89(11), 945-965.

Hutchinson, J. W. (1990). Mixed mode fracture mechanics of interfaces.
Metal-ceramic interfaces, 295-306.

Hutchinson, J. W. \& Suo, Z. 1992 Mixed mode cracking in layered materials. In
Advances in applied mechanics, vol. 29 (eds J. W. Hutchinson \& T. Y. Wu), pp.
63--191. Boston, MA: Academic Press.

\bigskip

Johnson KL, 1996, Continuum mechanics modeling of adhesion and friction.
Langmuir 12:4510--4513.

Johnson, K. L., 1997, Adhesion and friction between a smooth elastic spherical
asperity and a plane surface. In Proceedings of the Royal Society of London
A453, No. 1956, pp. 163-179).

Johnson, K. L., Kendall, K. \& Roberts, A. D. 1971 Surface energy and the
contact of elastic solids. Proc. R. Soc. Lond. A 324, 301--313.

Maugis, D. (2013). Contact, adhesion and rupture of elastic solids (Vol. 130).
Springer Science \& Business Media.

\bigskip Menga, N., Carbone, G., \& Dini, D. (2018). Do uniform tangential
interfacial stresses enhance adhesion?. Journal of the Mechanics and Physics
of Solids, 112, 145-156.

Nguyen, D. T., Paolino, P., Audry, M. C., Chateauminois, A., Fretigny, C., Le
Chenadec, Y., ... \& Barthel, E. (2011). Surface pressure and shear stress
fields within a frictional contact on rubber. The Journal of Adhesion, 87(3), 235-250.

Papangelo, A., Scheibert, J., Sahli, R., Pallares, G., \& Ciavarella, M.
(2019). Shear-induced contact area anisotropy explained by a fracture
mechanics model. Physical Review E, 99(5), 053005.

Papangelo, A., \& Ciavarella, M. (2019). On mixed-mode fracture mechanics
models for contact area reduction under shear load in soft materials. Journal
of the Mechanics and Physics of Solids, 124, 159-171.

Sahli, R., Pallares, G., Papangelo, A., Ciavarella, M., Ducottet, C., Ponthus,
N., \& Scheibert, J. (2019). Shear-induced anisotropy in rough elastomer
contact. Physical Review Letters, 122(21), 214301.

Savkoor, A. R. \& Briggs, G. A. D. 1977 The effect of a tangential force on
the contact of elastic solids in adhesion. Proc. R. Soc. Lond. A 356, 103--114.

Sahli, R. Pallares, G. , Ducottet, C., Ben Ali, I. E. , Akhrass, S. Al ,
Guibert, M. , Scheibert J. , Evolution of real contact area under shear,
Proceedings of the National Academy of Sciences, 2018, 115 (3) 471-476; DOI: 10.1073/pnas.1706434115

Tabor, D. (1977) Surface forces and surface interactions. J. Colloid Interface
Sci. 58, 2.

Waters JF, Guduru PR, 2009, Mode-mixity-dependent adhesive contact of a sphere
on a plane surface. Proc R Soc A 466:1303--1325.

\end{document}